\documentclass[10pt,preprint]{aastex}
\usepackage[dvips]{color}
\usepackage{graphicx}
\usepackage{enumerate}

\def\msol{\hbox{\kern 0.20em $M_\odot$}}
\def\lsol{\hbox{\kern 0.20em $L_\odot$}}
\def\rsol{\hbox{\kern 0.20em $R_\odot$}}
\def\sr{\hbox{\kern 0.20em sr}}
\def\srmu{\hbox{\kern 0.20em sr$^{-1}$}}
 
\def\g{\hbox{\kern 0.20em g}}
\def\gmu{\hbox{\kern 0.20em g$^{-1}$}}
\def\kg{\hbox{\kern 0.20em kg}}
\def\pc{\hbox{\kern 0.20em pc}}
 
\def\mum{\hbox{\kern 0.20em $\mu$m}}
\def\mumd{\hbox{\kern 0.20em $\mu$m$^{-2}$}}
\def\cm{\hbox{\kern 0.20em cm}}
\def\m{\hbox{\kern 0.20em m}}
\def\km{\hbox{\kern 0.20em km}}
\def\nm{\hbox{\kern 0.20em nm}}
 
\def\s{\hbox{\kern 0.20em s}}
\def\h{\hbox{\kern 0.20em h}}
\def\sec{\hbox{\kern 0.20em sec}}
\def\min{\hbox {\kern 0.20em min}}
\def\smu{\hbox{\kern 0.20em s$^{-1}$}}
\def\smd{\hbox{\kern 0.20em s$^{-2}$}}
\def\an{\hbox{\kern 0.20em an}}
\def\anmu{\hbox{\kern 0.20em an$^{-1}$}}
\def\deg{\hbox{\kern 0.20em $^{\rm o}$}}
\def\yr{\hbox{\kern 0.20em yr}}
\def\yrmu{\hbox{\kern 0.20em yr$^{-1}$}}
\def\Myr{\hbox{\kern 0.20em Myr}}
\def\Mymu{\hbox{\kern 0.20em Myr$^{-1}$}}
\def\K{\hbox{\kern 0.20em K}}
\def\pcmu{\hbox{\kern 0.20em pc$^{-1}$}}
\def\pcmd{\hbox{\kern 0.20em pc$^{-2}$}}
\def\pcmt{\hbox{\kern 0.20em pc$^{-3}$}}
\def\kms{\hbox{\kern 0.20em km\kern 0.20em s$^{-1}$}}
\def\kmpd{\hbox{\kern 0.20em km$^{2}$}}
\def\kpc{\hbox{\kern 0.20em kpc}}
\def\cms{\hbox{\kern 0.20em cm\kern 0.20em s$^{-1}$}}
\def\erg{\hbox{\kern 0.20em erg}}
\def\ergs{\hbox{\kern 0.20em erg}}
\def\cmpd{\hbox{\kern 0.20em cm$^2$}}
\def\cmmd{\hbox{\kern 0.20em cm$^{-2}$}}
\def\cmms{\hbox{\kern 0.20em cm$^{-6}$}}
\def\cmpt{\hbox{\kern 0.20em cm$^3$}}
\def\cmmt{\hbox{\kern 0.20em cm$^{-3}$}}
\def\mpd{\hbox{\kern 0.20em m$^2$}}
\def\mmd{\hbox{\kern 0.20em m$^{-2}$}}
\def\mpt{\hbox{\kern 0.20em m$^3$}}
\def\mmt{\hbox{\kern 0.20em m$^{-3}$}}
\def\mujy{\hbox{\kern 0.20em $\mu$Jy}}
\def\mjy{\hbox{\kern 0.20em mJy}}
\def\Mj{\hbox{\kern 0.20em MJy}}
\def\jy{\hbox{\kern 0.20em Jy}}
\def\ghz{\hbox{\kern 0.20em GHz}}
\def\srmd{\hbox{\kern 0.20em sr$^{-1}$}}
\def \kms{km~$\rm{s}^{-1}$}

\def \mum{$\mu$m}

\def\G{\hbox{\kern 0.20em G}}

\def\h13cop{\hbox{H$^{13}$CO$^{+}$}}

\def\S+{\hbox{S{\small II}}}

\slugcomment{The Evolving ISM in the Milky Way \& Nearby Galaxies}

\shorttitle{FIR and Radio Morphologies of Galaxy Disks}
\shortauthors{Murphy et al.}

\begin{document}

\newcommand{\jfourteen}{\hbox{$J=14\rightarrow 13$}}
 \title{Learning about the Recent Star Formation History of Galaxy Disks by Comparing their Far-Infrared and Radio Morphologies: Cosmic-Ray Electron Diffusion after Star Formation Episodes}

\author{E.J.~Murphy,\altaffilmark{1} G.~Helou,\altaffilmark{2}, J.D.P.~Kenney,\altaffilmark{1} L~Armus,\altaffilmark{3} and R~Braun\altaffilmark{4}
}

\altaffiltext{1}{\scriptsize Department of Astronomy, Yale University,
  P.O. Box 208101, New Haven, CT 06520-8101; murphy@astro.yale.edu}
\altaffiltext{2}{\scriptsize California Institute of Technology, MC
  314-6, Pasadena, CA 91125}
\altaffiltext{3}{\scriptsize {\it Spitzer} Science Center, California
  Institute of Technology, Pasadena, CA 91125} 
\altaffiltext{4}{\scriptsize ASTRON, P.O. Box 2, 7990 AA Dwingeloo,
  The Netherlands}

\begin{abstract}
We present results on the interstellar medium (ISM) properties of 29
galaxies based on a comparison of {\it Spitzer} far-infrared and
Westerbork Synthesis Radio Telescope radio continuum imagery.
Of these 29 galaxies, 18 are close enough to resolve at $\la$1~kpc
scales at 70~$\micron$ and 22~cm.    
We extend the \citet{ejm06a,ejm06b} approach of smoothing infrared
images to approximate cosmic-ray (CR) electron spreading and thus
largely reproduce the appearance of radio images. 

Using a wavelet analysis we decompose each 70~$\micron$ image into one
component containing the star-forming {\it structures} and a second
one for the diffuse {\it disk}.   
The components are smoothed separately, and their combination compared
to a free-free corrected 22~cm radio image; 
the scale-lengths are then varied to best match the radio and smoothed
infrared images.  
We find that late-type spirals having high amounts of ongoing star
formation benefit most from the two-component method.  
We also find that the disk component dominates 
for galaxies having low star formation activity, whereas
the structure component dominates at high star formation activity.  

We propose that this result arises from an age effect rather than from
differences in CR electron diffusion due to varying ISM parameters.
The bulk of the CR electron population in actively star-forming galaxies
is significantly younger than that in less active galaxies due to recent
episodes of enhanced star formation; 
these galaxies are observed within $\sim$10$^{8}$~yr since the onset of the most recent star formation episode.
The sample irregulars have anomalously low best-fit scale-lengths for
their surface brightnesses compared to the rest of the sample spirals
which we attribute to enhanced CR electron escape.    
\end{abstract}

  
\lefthead{Murphy et al.}
\righthead{FIR and Radio Morphologies of Galaxy Disks}

\section{Introduction}

To date, most of our knowledge about the relativistic phase of the interstellar medium (ISM) outside of the Galaxy, consisting of relativistic charged particles and magnetic fields, has been obtained indirectly through the detection of synchrotron emission via multi-frequency radio observations \citep[e.g.][]{nd91,dlg95,ul96,ji99,rb05}.
Synchrotron emission arises from cosmic-ray (CR) electron energy losses as these
particles are accelerated in the magnetic fields of galaxies.
Although the energy density in CR electrons is only $\sim$1\% of that
for CR nuclei, the similarity between the spatial distributions of
gamma-ray and synchrotron emission within the Galaxy suggests that CR
electrons and CR nuclei are fairly well mixed on the scales of a few
hundred parsecs \citep[e.g.][]{ch82,jb86,ww91}.
The spatial distribution of a galaxy's synchrotron emission is a
function of a galaxy's CR electron and magnetic field distributions. 
Thus, radio synchrotron maps provide only limited insight on the
source distribution of the CR electrons as well as the distances the
particles may have traveled before ending up in their current location
of emission.

Massive stars ($\ga$8~$M_{\sun}$) are the progenitors of supernovae
(SNe) whose remnants (SNRs), through the process of diffusive shock
acceleration \citep{ab78,bo78}, appear to be the main acceleration
sites of CR electrons responsible for a galaxy's observed synchrotron
emission. 
These same young massive stars are often the primary sources for dust
heating as they emit photons which are re-radiated at far-infrared
(FIR) wavelengths. 
This shared origin between the FIR and radio emission of galaxies is
thought to be the foundation for the observed FIR-radio correlation
among 
\citep[e.g.][]{de85,gxh85,sn97,nb97,yun01}
and within galaxies 
\citep[e.g.][]{bg88,Xu92,mh95,hoer98,hip03,ejm06a,ah06}.

Coupling the shared origin of a galaxy's FIR and radio emission
with the fact that the mean free path of dust-heating photons
($\sim$100~pc) is significantly shorter than the expected diffusion
length of CR electrons ($\sim$1-2~kpc) led \citet{bh90} to conjecture
that the radio image of a galaxy should resemble a smoothed version of
its infrared image.  
Consequently, it appears that the close spatial correlation between
the FIR and radio continuum emission within galaxies can be used to
characterize the propagation history of CR electrons.  
This prescription has been shown to hold for galaxies observed at the
``super resolution'' ($\la$1$\arcmin$) of {\it IRAS} HIRES data
\citep{mh98} and, more recently, for high resolution
($\sim$18$\arcsec$) {\it Spitzer} 70$~\micron$ imaging 
\citep[][hereafter M06a]{ejm06a}.
This phenomenology has been further corroborated on
scales $\ga$50~pc by \citet{ah06} who find synchrotron haloes around
individual star-forming regions are more extended than FIR-emitting
regions within the Large Magellanic Cloud.

\citet[][hereafter, M06b]{ejm06b} recently studied how the spatial
distributions of a galaxy's FIR and radio emission vary as a function
of the intensity of star formation. 
They concluded that CR electrons are, on average, younger and closer
to their place of origin within galaxies having higher amounts of star
formation activity compared with more quiescent galaxies.
Using a wavelet-based image decomposition, we extend this work by
attempting to characterize separately
CR electron populations associated with a galaxy's diffuse disk and
its star-forming complexes.  
We carry out this study for a sample of galaxies observed as part of
the {\it Spitzer} Infrared Nearby Galaxies Survey
\citep[SINGS;][]{rk03} and the  Westerbork Synthesis Radio Telescope
\citep[WSRT-SINGS;][]{rb07} for which we have the spatial resolution
to resolve physical scales $<1$~kpc.
The complete study can be found in \citet{ejm08}.

\section{Observations and Analysis}
Observations at 24, 70, and 160~$\micron$ were obtained using the
Multiband Imaging Photometer for {\it Spitzer} \citep[MIPS;][]{gr04}
as part of the SINGS legacy science program.  
A detailed description of the SINGS observational strategy can be
found in \citet{rk03}.  
Radio continuum imaging at 22~cm was performed using the Westerbork
Synthesis Radio Telescope (WSRT) as part of the WSRT-SINGS survey.
A complete description of the radio observations and image processing
steps can be found in \citet{rb07}.  
We match the resolution of the MIPS and radio images using Gaussian
PSFs rather than the MIPS PSFs, which suffer from significant power in
their side-lobes. 

We compute the radiation field energy density ($U_{\rm rad}$) of
each galaxy using its TIR ($3-1100~\micron$) surface brightness since this parameter is sensitive to the diffusion of CR electrons.
Using the deprojected area of elliptical apertures ($A_{\rm TIR}$ ), we calculate TIR surface
brightnesses $(\Sigma_{\rm TIR} = L_{\rm TIR}/A_{\rm TIR})$ along with
estimates of $U_{\rm rad}$ for radiation emitted near the surface
of a semi-transparent body such that
\begin{equation}
\label{eq-urad}
U_{\rm rad} \approx \frac{2\pi}{c}I_{\rm bol} \ga \frac{L_{\rm
    TIR}}{2A_{\rm TIR}c}\left(1 + \sqrt{\frac{3.8 \times
    10^{42}}{L_{\rm TIR}}}\right), 
\end{equation}
where $I_{\rm bol}$ is a galaxy's bolometric surface brightness, $c$
is the speed of light, and all quantities are given in cgs units.
The parenthetical term in Equation \ref{eq-urad} provides a correction
for non-absorbed UV emission that was empirically derived by
\citet{efb03} using archived FIR and UV data for a sample of more than
200 galaxies.

\subsection{Multi-Component Image-Smearing Model}
Using a phenomenological image-smearing model, first presented by
\citet{bh90}, M06b studied how the spatial distributions of FIR and
radio emission varied as a function of star-formation intensity within
12 spiral galaxy disks.
The basic procedure used here is similar to that presented in M06a,b.
We calculate and minimize the residuals between the free-free corrected radio and
observed infrared images after convolving the infrared maps by a 
parameterized kernel  $\kappa({\bf r})$.
The new element is that we now, using a wavlet-based technique, decompose each observed infrared image, into two sub-images:
(1) a {\it structure} image containing features with spatial scale smaller than 1~kpc; (2) a {\it disk} image containing all structures with characteristic spatial scales larger than or equal to
1~kpc, largely constituting a galaxy's diffuse disk.  
The {\it structure} and {\it disk} images are smeared independently, summed, and then compared to the radio image.
For a full description of the phenomenological model we refer the reader to \citet{ejm08}.

\begin{figure}[!ht]
\centerline{\hbox{
  \resizebox{16cm}{!}{
\plottwo{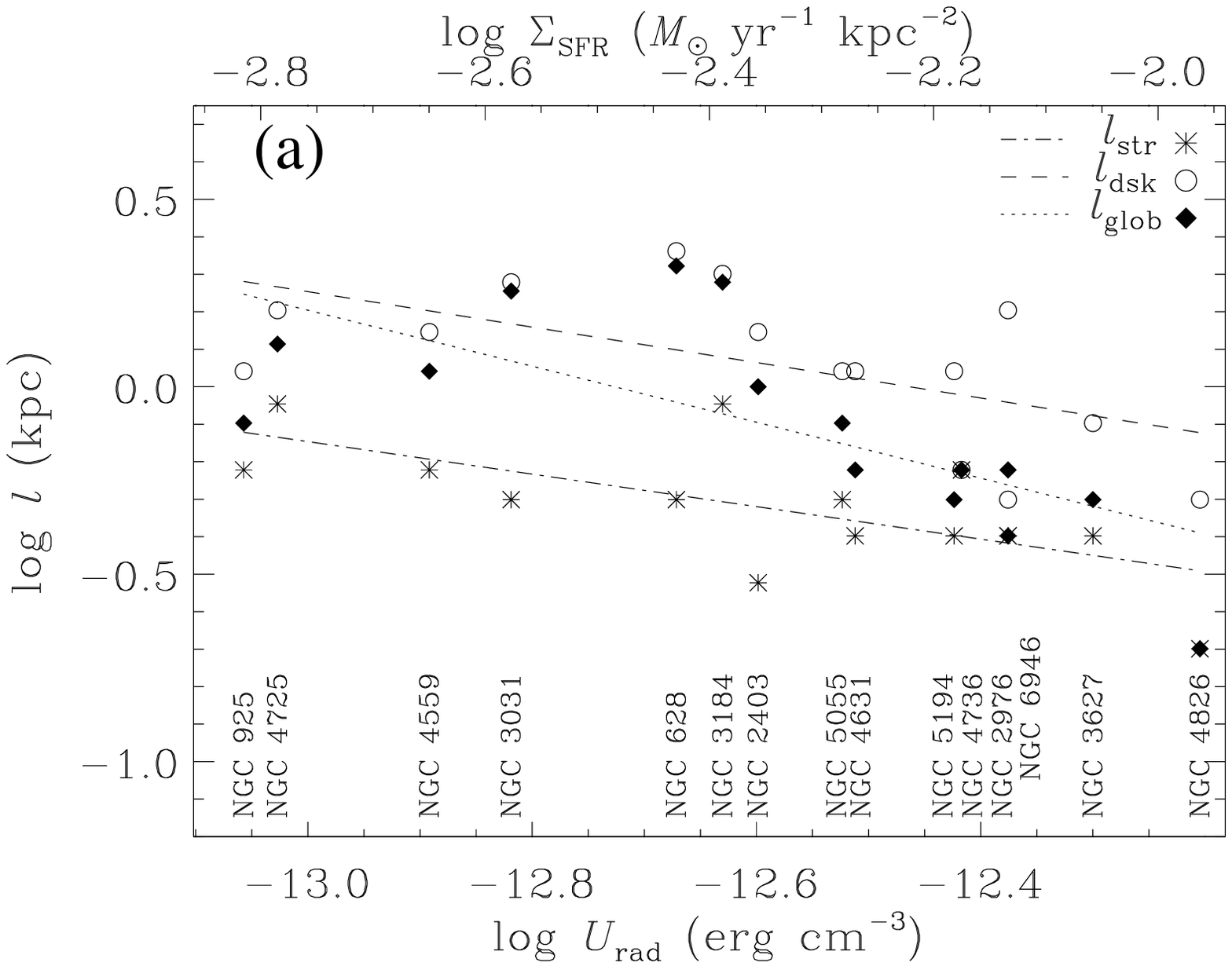}{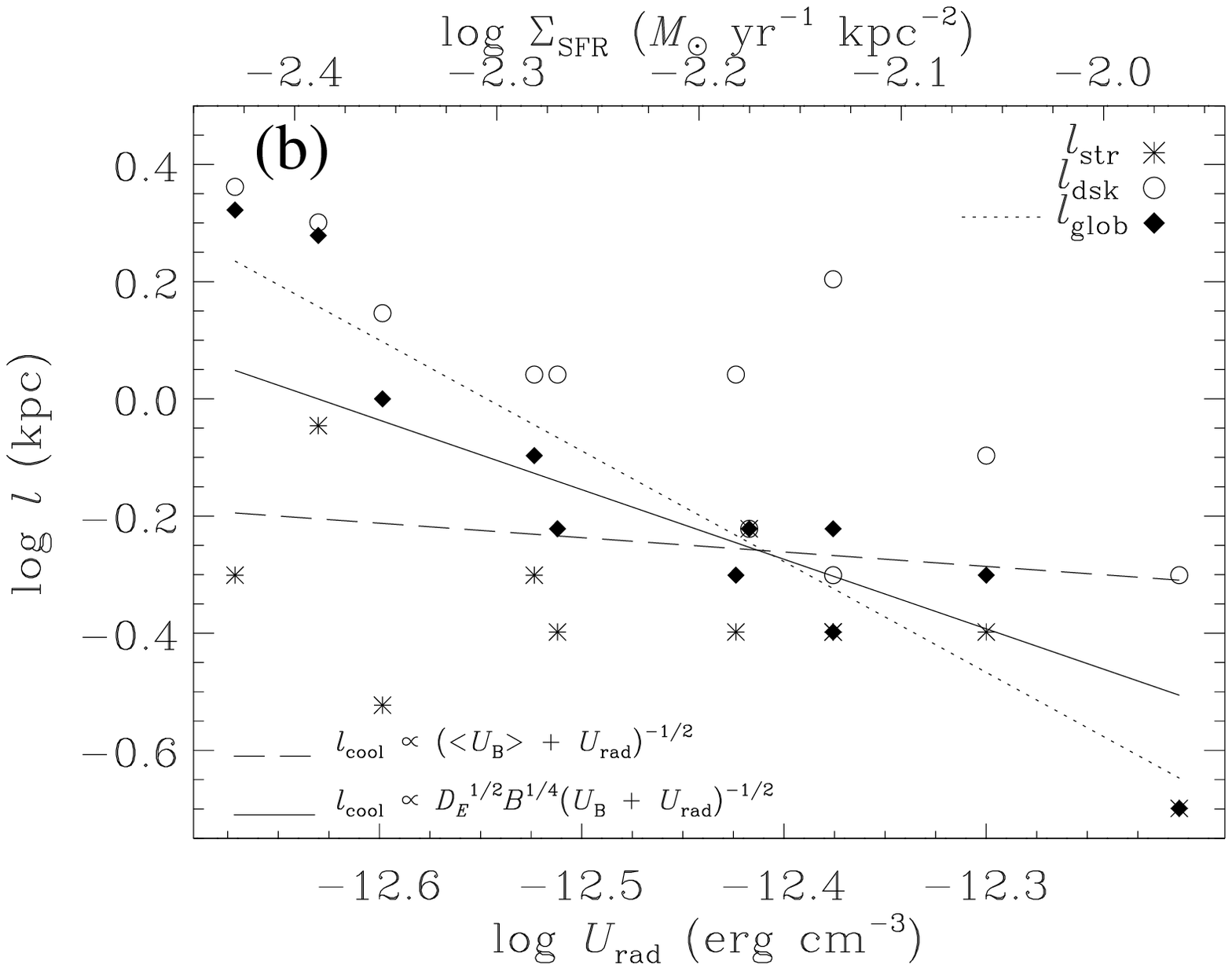}}}}
\caption{\footnotesize
    In panel (a) we plot the best-fit disk, structure, and global 
    scale-lengths for galaxies in our sample which are resolved at
    scales less than 1~kpc against the radiation field energy
    densities, $U_{\rm rad}$.
    We exclude the galaxies Holmb~II, IC~2574, and NGC~4236 which
    have morphologies not well fit by our phenomenological
    model.
    Least square fits for the best-fit disk, structure, and global
    scale-lengths are plotted as {\it dot-dash}, {\it dashed}, and
    {\it dotted lines}, respectively. 
    In panel (b) we have excluded the galaxies NGC~925, NGC~4725,
    NGC~4559, and NGC~3031 due to possible signal-to-noise effects.
    Along with the fit to the global scale-lengths ({\it dotted line})
    we also plot with the expected diffusion scale-lengths due to
    inverse Compton (IC) losses in a fixed magnetic field ({\it
    long-dashed line}) and synchrotron + IC losses with an
    energy-dependent diffusion coefficient $D_{E}$ for the steepest
    possible index ({\it solid line}).
    The slope of our observed trend is clearly steeper than what is achieved by our scaling-relations;  this suggests that the variation of ISM parameters alone cannot explain out observations.     
  \label{fig-1}}
\end{figure}

\section{Discussion}
By plotting the smearing scale-length which best matches the infrared image to the observed radio map versus TIR surface brightness (see Figure \ref{fig-1}), we find that the FIR and non-thermal radio morphologies are more similar to each other for galaxies having higher radiation field energy densities compared to galaxies with lower radiation field energy densities.  
Following the interpretation of M06b, our results indicate that CR
electrons are, on average, closer to their place of origin in galaxies
having higher star formation activity.

As CR electrons propagate through the ISM of galaxies they lose their
energy due to a number physical processes including synchrotron, inverse-Compton (IC) scattering,
bremsstrahlung, ionization, and adiabatic expansion losses.
In normal galaxies synchrotron and IC scattering processes are likely the most significant energy loss terms for CR electrons associated with 1~GHz emission \citep{jc92}; 
the other terms listed will become non-negligible, however, for
galaxies hosting extreme episodes of star-formation like starbursting
ultra-luminous infrared galaxies (ULIRGs) \citep[e.g.][]{tt06}.  

A CR electron having energy $E$ will emit most of its energy at a critical frequency $\nu_{\rm c}$ where  
\begin{equation} 
\label{eq-nuBE}
 \left(\frac{\nu_{\rm c}}{\rm GHz}\right) = 1.3\times10^{-2}
  \left(\frac{B}{\rm \mu G}\right)
  \left(\frac{E}{\rm GeV}\right)^{2}.
\end{equation}
From this, we can express the effective cooling timescale for CR electrons due to synchrotron
and IC losses as 
\begin{equation}
\label{eq-cool}
  \left(\frac{\tau_{\rm cool}}{\rm yr}\right) \sim 5.7\times10^{7}
  \left(\frac{\nu_{\rm c}}{\rm GHz}\right) ^{-1/2}
  \left(\frac{B}{\rm \mu G}\right)^{1/2}
  \left(\frac{U_{\rm B}+U_{\rm rad}}{10^{-12}~{\rm
  erg~cm^{-3}}}\right)^{-1}.
\end{equation}
In simple diffusion models, the propagation of CR electrons is
usually characterized by an empirical, energy-dependent diffusion
coefficient, $D_{E}$ \citep[e.g.][]{ginz80}.
The value of $D_{E}$ has been found to be around \(4-6\times
10^{28}~{\rm cm^{2}~s^{-1}}\) for $\la$GeV CRs by fitting diffusion
models with direct measurements of CR nuclei 
(i.e. secondary-to-primary ratios like Boron-to-Carbon) 
within the Solar Neighborhood \citep[e.g.][]{im02}.  
We will assume,
\begin{equation}
\label{eq-DE}
\left(\frac{D_{E}}{\rm cm^{2}~s^{-1}}\right) \sim \Bigg\{
\begin{array}{cc}
5 \times 10^{28}, & E < 1~{\rm GeV}\\
5 \times 10^{28}(\frac{E}{{\rm GeV}})^{1/2}, & E \geq 1~{\rm GeV}.
\end{array}
\end{equation}

Now, neglecting escape and using a simple random-walk equation, CR electrons will 
diffuse a distance \(l_{\rm cool} = (D_{E}\tau_{\rm cool})^{1/2}\)
before losing all of their energy to synchrotron and IC losses.
By combining Equations \ref{eq-nuBE} and \ref{eq-DE}, we can express
$D_{E}$ as a function of $B$ for a fixed $\nu_{\rm c}$ such that, for
CR electrons having energies $\geq$1~{\rm GeV},
\begin{equation} 
\label{eq-diff}
\left(\frac{l_{\rm cool}}{\rm kpc}\right) \sim 7 \times 10^{-4} 
\left(\frac{\tau_{\rm cool}}{\rm yr}\right)^{1/2}
\left(\frac{\nu_{\rm c}}{\rm GHz}\right)^{1/8}
\left(\frac{B}{\rm \mu G}\right)^{-1/8}.
\end{equation}

\subsection{Order of Magnitude Estimates}
Using the above equations, we derive simple, order-of-magnitude estimates to determine whether diffusion and cooling of CR electrons in a steady-state star formation model are able to account for our observations.
Taking the mean value of $U_{\rm rad}$ for those galaxies plotted in
Figure \ref{fig-1}b (i.e. $3.7\times 10^{-13}$erg~cm$^{-3}$), and
assuming $U_{\rm rad}=U_{\rm B}$, which has been shown to be a
reasonable assumption for a large sample of spiral galaxies
\citep{lvx96}, we find from Equation \ref{eq-cool} that the average
cooling time for a 1.4~GHz emitting CR electron is $\sim$$1.1\times
10^{8}$~yr.  
Inserting this value into Equation \ref{eq-diff}, we measure a
diffusion scale-length of $\sim$6.8~kpc; this value is clearly off of
the scale shown in Figure \ref{fig-1}b.  
On the other hand, if we instead assume a fixed, typical magnetic field 
strength of 9~$\mu$G \citep{sn95}, and that $U_{\rm B}=U_{\rm rad}=3.2\times
10^{-12}$~erg~cm$^{-3}$, the average cooling time for a 1.4~GHz emitting
CR electron is $\sim$$2.2\times 10^{7}$~yr with a diffusion
scale-length of $\sim$2.6~kpc.  
While this value for $U_{\rm rad}$ is much higher than what we infer
from the average TIR surface brightness of the sample, it must apply
near bright star-forming structures, whose TIR surface brightnesses
are much greater.  
Even so, this diffusion scale-length is much larger than any value we
find for the best-fit structure scale-lengths.
From these simple order of magnitude estimates it appears that
particle fading due to cooling by Inverse Compton and synchrotron
processes alone cannot explain the structural differences between the
FIR and radio maps; 
the best explanation seems to be differences in the CR electron population ages.

\begin{figure}[!ht]
\centerline{\hbox{
  \resizebox{9cm}{!}{
\plotone{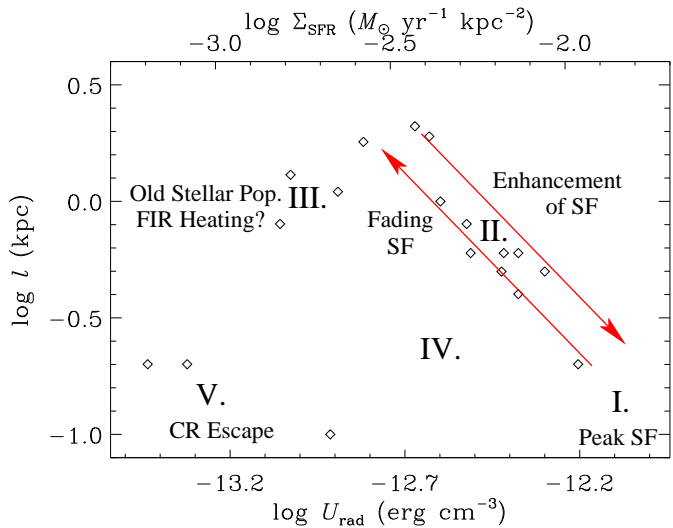}}}}
\caption{\footnotesize
The best-fit global scale-lengths for the entire 18 galaxy
  sample using isotropic kernels having an exponential profile and our
  free-free corrected radio maps.  
  Overplotted on the scatter diagram are roman numerals to identify
  the general placement of galaxies on this diagram; these locations are discussed in the text.
  \label{fig-2}}
\end{figure}

\subsection{Morphologies of Star Formation}
In order to provide a physical interpretation for the location of galaxies in the best-fit scale-length---$U_{\rm rad}$ diagram, we describe their general locations, designated by roman numerals, in Figure \ref{fig-2}.
While the placement of NGC~925, NGC~4725, NGC~4559 in Figure \ref{fig-1}a may be the result of a signal-to-noise effect, we will now consider alternative physical explanations.
The irregular galaxies in our sample (Holmb~II, IC~2574, and NGC~4236) have also been excluded thus far in our discussion due to their lack of a disk component in the FIR and radio.  
The placement of these objects in Figure \ref{fig-2} and their behavior will now be put into the context of our phenomenological picture.

\begin{enumerate}[I.]

\item 
  In this region of the diagram we assume a galaxy has just reached a
  peak in its surface brightness after a recent episode of enhanced
  star formation; 
  the episode must have occurred within the last few Myr to ensure that
  SNRs are still relatively young and that the morphologies of the 
  radio and infrared emission are similar because CR electrons have
  not had time to diffuse very far.

\item
 After reaching a peak in global star formation activity, those CR
  electrons associated with the recent enhancement of star formation
  will dominate the CR population.
  As they spread through the galaxy they will begin to lose their
  energy to synchrotron and IC processes. 
  A galaxy's position along this trend cannot be due to variations in ISM parameters alone; rather, its star formation history is the dominant parameter (see Figure \ref{fig-1}b).
  Galaxies may move along this part of phase space as star formation activity becomes enhanced and then fades.

\item 
 Galaxies in this region are characterized by low star formation activity and have best-fit scale-lengths which are shorter than expected given the trend found in region II.  
While this situation can be explained by a signal-to-noise effect, it is can also be explained by a significant amount of FIR emission arising from an old stellar population.  

\item
We find no galaxies occupying this region of phase space in which relatively short global scale-lengths would be measured for a moderate values of $U_{\rm rad}$.  
  One way for a galaxy to populate this part of the diagram is if, after a long period of quiescence, it begins to form stars at a moderate rate.  
  The galaxy would then pass through this region very quickly ($\la$10$^{7}$~yr) before shifting into region II.  
  The lack of such galaxies suggests that, at least in the local Universe, star formation in spirals does not completely cease for long periods of time.

\item
  The irregulars behave markedly different than the sample spirals.  
  We do not believe this discrepant behavior to be the result of
  signal-to-noise effects.
The morphologies and ISM of these galaxies do not seem consistent with
  keeping their CR electrons bound outside of their initial clouds around SNRs.  
  After the CR electrons leave this cloud, the lack of a dense ISM and
  magnetic field to keep them trapped through multiple scatterings off
  of magneto-hydrodynamic (MHD) waves allows them to easily escape the
  system and enter intergalactic space.

\end{enumerate}

\section{Summary and Conclusions \label{sec-conc}}
Using a two-component image-smearing analysis, we have separated the
signatures of CR electron diffusion at spatial scales corresponding to
star-forming structures ($<$1~kpc) and galaxy disks ($\geq$1~kpc) 
within 18 galaxies observed as part of SINGS and WSRT-SINGS. 
Our results and conclusions can be summarized as follows:
\begin{enumerate}

  
\item 
  The best-fit global scale-lengths decrease as a function of increasing star
  formation activity as measured by the infrared surface brightness of
  a galaxy.  
  Our interpretation is that a galaxy's CR electrons are closer
  to their place of origin within galaxies having intense star formation
  activity. 

\item
  The trend of decreasing best-fit {\it global} scale-length with
  increasing radiation field energy density is due to higher surface
  brightness galaxies having undergone a recent enhancement of star
  formation rather than variations in other ISM parameters.    
For sufficiently large enhancements, these galaxies are observed within $\sim$10$^{8}$~yr of the onset of the most recent star formation episode.

\item
  Unlike spirals, irregular galaxies lack any well defined diffuse disk
  component at either 70~$\micron$, or especially at 22~cm.
  Presumably, the CR electrons escape these galaxies soon after
  leaving their parent star-forming regions due the absence of a
  dense ISM which would keep large-scale interstellar magnetic field
  locked into place.
  This conclusion helps to explain why these galaxies have global
  FIR/radio ratios systematically greater than canonical values.

\item 
As infrared surface brightness increases, the characteristic diffusion scale-length of a galaxy's CR electron population begins to transition at \(\log~U_{\rm rad} \leq -12.5\) from being biased by CR electrons making up its diffuse disk to being biased by those recently injected near star-forming structures.
  From this we conclude that a galaxy's CR electron population 
  transitions from being dominated by old CR electrons to
  being dominated by young CR electrons as a function of star
  formation intensity.

\item
 The two-component analysis works better than smearing with a single smoothing kernel for spiral galaxies of type Sb or later which have high amounts of ongoing star formation activity (i.e. $\sim$40\% of the sample).
  This result suggests that star formation must be intense and highly structured for the two-component analysis of these data to differentiate properly between the different CR electron populations.

\end{enumerate}

\acknowledgements
E.J.M would also like to thank other members of the {\it Spitzer}
Infrared Nearby Galaxies Survey (SINGS) team for their invaluable
contributions to the work presented here, especially those of
R.C.Kennicutt, Jr., D.Calzetti, K.D.Gordon, C.W.Engelbracht, and
G.Bendo. 
As part of the {\it Spitzer} Space Telescope Legacy Science Program,
support was provided by NASA through Contract Number 1224769 issued by
the Jet Propulsion Laboratory, California Institute of Technology
under NASA contract 1407.  



\end{document}